\documentclass[a4paper,11pt]{scrartcl}
\usepackage[utf8x]{inputenc}
\usepackage{tikz}
\usepackage{hyperref}
\usepackage{amsmath}
\title{Rationality is Self-Defeating in Permissionless Systems%
\footnote{First posted on September 23, 2019 at \url{https://bford.info/2019/09/23/rational/}}}
\author{Bryan Ford%
\footnote{\'Ecole Polytechnique F\'ed\'erale de Lausanne, Switzerland, \url{bryan.ford@epfl.ch}}
\and 
Rainer Böhme%
\footnote{Universität Innsbruck, Austria, \url{rainer.boehme@uibk.ac.at}}
}

\begin{document}
\maketitle
\newcommand{\participant}{\ensuremath{P}}
\newcommand{\Sys}{\ensuremath{S}}
\newcommand{\Sprime}{\ensuremath{S'}}
\section{Introduction}

Many blockchain and cryptocurrency fans seem to prefer building and analyzing decentralized systems in a rational or ``greedy behavior'' failure model, rather than a Byzantine or ``arbitrary behavior'' failure model.
Many of the same blockchain and cryptocurrency fans also like open, permissionless systems like Bitcoin and Ethereum, which anyone can join and participate in using weak identities such as anonymous cryptography key pairs.

What most of these heavily-overlapping sets of fans do not seem to realize, however, is that rationality assumptions are self-defeating in open permissionless systems with weak identities.
A fairly simple metacircular argument---a kind of ``Gödel's incompleteness theorem \cite{Goedel1931} for rationality''---shows that for any system \Sys{} that makes \emph{any} behavioral assumption, including but not limited to a rationality assumption, a rational attacker both exists and \emph{has an incentive} to defeat that behavioral assumption, thereby violating that assumption and exhibiting Byzantine behavior from the perspective of the system.

As a quick summary of the argument we will expand below, suppose a permissionless system like Bitcoin is secure against rational attacks, but has some weakness against irrational Byzantine attacks in which the attacker would lose money.
Because the system is open, permissionless, and exists within a larger ecosystem, a rational attacker can find ways to ``bet against'' Bitcoin's security in \emph{other} financially-connected systems (e.\,g., Ethereum), making a profit \emph{outside of} Bitcoin on this attack against Bitcoin. An attack that appears irrational in the context of Bitcoin may be perfectly rational in the context of the larger ecosystem.

For this reason, an open permissionless system designed to be secure only against rational adversaries is actually just \emph{insecure}, unless it remains secure even when the ``rational'' participants become fully Byzantine. Given this, one might as well have designed the permissionless system in a Byzantine model in the first place. The rationality assumption offers no actual benefit, but merely can make an insecure system appear secure under flawed analysis.

This manuscript is based partly on ideas in the second author's lecture~\cite{Boehme2019} at the BDLT Summer School in Vienna \cite{BDLT2019}. While formalizing the argument would require some effort, we thought it would be worth at least sketching the argument intuitively for the public record.

\section{Threat Modeling: Honest, Byzantine, and Rational Participants}

In designing or analyzing the security of any decentralized system, we must define the system's \emph{threat model}, and in particular our assumptions about the behaviors of the participants in the system. 
An \emph{honest}, \emph{correct}, or \emph{altruistic} participant is one that we assume to follow the system's protocol rules as specified, hence representing a ``well-behaved'' participant exhibiting no adversarial behavior.

A \emph{Byzantine} participant, named after the Byzantine Generals Problem~\cite{Lamport1982}, is one we make \emph{no} assumptions about. A Byzantine participant can behave in \emph{arbitrary} fashion, without restriction, and hence by definition represents the strongest possible adversary.

We would like to build systems that could withstand \emph{all} participants being Byzantine, but this appears fundamentally impossible. 
We therefore in practice have to make threshold security assumptions, such as that over two-thirds of the participants in classic Byzantine consensus protocols are honest, or that the participants controlling over half the hash power in Bitcoin are well-behaved.

Even with threshold assumptions, however, building systems that resist Byzantine behavior is extremely difficult, and the resulting systems are often much more complex and inefficient than systems tolerating weaker adversaries. 
We may therefore be tempted to improve a design's simplicity or efficiency by making stronger assumptions about the behavior of adversarial participants, effectively weakening the assumed adversary.

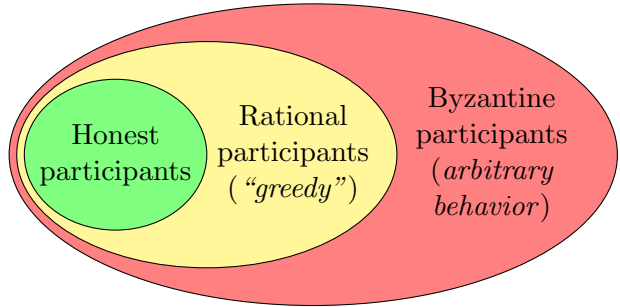
\begin{figure}
\begin{center}
\begin{tikzpicture}[>=stealth,y=2cm]
	\draw [draw,fill=red!50] (0,0) ellipse (4 and 1) node [right=2.5em,black] {\parbox{25mm}{\centering Byzantine\\participants\\(\emph{arbitrary behavior})}};
	\draw [draw,fill=yellow!50] (-1.4,0) ellipse (2.5 and .75) node [right,black] {\parbox{20mm}{\centering Rational\\participants\\ (\emph{``greedy''})}};
	\draw [draw,fill=green!50] (-2.6,0) ellipse (1.2 and .5) node [black] {\parbox{20mm}{\centering Honest\\participants}};
\end{tikzpicture}
\end{center}
\caption{Euler diagram of possible behaviors in the three threat models considered}
\label{fig:adversaries-1}
\end{figure}

One such popular assumption, especially in economic circles, is \emph{rationality}. 
In essence, we assume that rational participants may deviate from the rules in arbitrary ways but \emph{only when doing so is in their economic self-interest}, improving their expected rewards---usually but not always financial---in comparison with following the rules honestly~(cf.~Fig.~\ref{fig:adversaries-1}).

By assuming that adversarial participants are rational rather than Byzantine, we need not secure the system against \emph{all} possible participant behaviors, such as against participants who pay money with no reward merely to sow chaos and destruction.
Instead, we merely need to prove that the system is \emph{incentive compatible}, for example by showing that its rules represent a Nash equilibrium, in which deviations from the equilibrium will not give participants a greater financial reward.

Besides simplicity and efficiency, another appeal of rationality assumptions is the promise of \emph{strengthening} the system's security by lowering the threshold of participants we assume to be fully honest. 
To circumvent the classical Byzantine consensus requirement that fewer than one third of participants may be faulty, for example, we might hope to tolerate closer to 50\,\%, or even 100\,\%,
of participants being ``adversarial'' if we assume they are rational and not Byzantine.
Work on the Byzantine-Altruistic-Rational (BAR) model~\cite{Aiyer2005} 
and $(k,t)$-robustness~\cite{Abraham2011} 
exemplifies this goal, which sometimes appears achievable in closed systems with strong identities. 
But a direct implication of our metacircular argument is that an \emph{open} system cannot generally be secure if all participants are either Byzantine or rational.

\section{Assumptions Underlying the Argument}

The metacircular argument makes three main assumptions.

\paragraph{Assumption~1} First, the system \Sys{} under consideration is open and permissionless, allowing anyone to join and participate in the system using only weak, anonymous identities such as bare cryptographic key pairs.
Identities in \Sys{} need not even be costless~\cite{Friedman2001} provided their price is modest: the argument still works even if \Sys{} imposes membership fees or requires new wallet keys to be ``mined,'' for example. 
Proof-of-Work cryptocurrencies such as Bitcoin and Ethereum, Proof-of-Stake systems such as Algorand~\cite{Chen2019} and Ouroboros~\cite{Kiayias2017}, and most other permissionless systems seem to satisfy this openness property. 
Because participation is open to anyone globally and can be anonymous, we cannot reasonably expect police or governments to protect \Sys{} from attack: even if they wanted to and considered it their job, they would not be able to find or discipline a smart rational attacker who might be attacking from anywhere around the globe, especially from a country with weak international agreements and extradition rules.
Thus, \Sys{} must ``stand on its own,'' by successfully either withstanding or disincentivizing attacks coming from anywhere. 
(And it will turn out that merely disincentivizing such attacks is impossible.)

\paragraph{Assumption~2} Second, the system \Sys{} does not control a majority of total economic power or value in the world: i.\,e., it is not totally economically dominant from a global perspective. 
Instead, there may be (and probably are) actors outside of \Sys{} who, if rationally incentivized to do so, can at least temporarily muster an amount of economic power outside of \Sys{} comparable to or greater than the economic value within or controlled by \Sys{}. In other words, we assume that \Sys{} is not the ``biggest fish in the ocean.''
Given that there can be at most one globally dominant economic system at a time, it seems neither useful nor advisable to design systems that are secure only when they are the biggest fish in the ocean, because almost always they are not.

\paragraph{Assumption~3} Third, the system \Sys{} actually \emph{leverages} in some fashion the behavioral assumption(s) it makes on participants, such as a rationality assumption. 
That is, we assume there exist one or more (arbitrary) behavioral strategies that \Sys{} assumes some participants \emph{will not} follow, such as economically-losing behaviors in the case of rationality.
Further, we assume there exists such an assumption-violating strategy that will cause \Sys{} to malfunction or otherwise deviate observably from its correct operation. In fact, we need not assume that this deviant behavior will \emph{always} succeed in breaking \Sys{}, but only that it will non-negligibly \emph{raise the probability} of \Sys{} failing.
If this were not the case, and \Sys{} in fact operates correctly, securely, and indistinguishably from its ideal
even if participants do violate their behavioral assumptions, then \Sys{} is actually Byzantine secure after all.
In that case, \Sys{} is not actually benefiting from its assumptions about participant behavior, which are redundant and thus may be simply discarded.

\section{The Metacircular Argument: Rational Attacks on Rationality}

Suppose permissionless system \Sys{} is launched, and operates smoothly for some time, with all participants conforming to \Sys{}'s assumptions about them. Because \Sys{} is permissionless (assumption~1) and exists in a larger open world (assumption~2), new rational participants may arrive at any time, attracted by \Sys{}'s success and presumably growing economic value provided there is an opportunity to profit from doing so.

Consider a particular newly-arriving participant \participant{}. \participant{} could of course play by the rules \Sys{} assumes of \participant{}, in which case the greatest immediate economic benefit \participant{} could derive from participating in \Sys{} is some fraction of the total economic value currently embodied in \Sys{} (e.\,g., its market cap). 
For most realistic permissionless systems embodying strong founders' or early-adopters' rewards, if \participant{} is not one of the original founders of \Sys{} but arrives substantially after launch, then \participant{}'s near-term payoff prospectives from joining \Sys{} is likely bounded to a fairly \emph{small} fraction of \Sys{}'s total value.
But what if there were another strategy \participant{} could take, for perfectly \emph{rational} and economically-motivated reasons, by which \participant{} could in relatively short order acquire a \emph{large} fraction of \Sys{}'s total value?

\begin{figure}
\begin{center}
\begin{tikzpicture}[>=stealth]

	\draw [gray!15,fill,rounded corners] (-5.5,-1.5) rectangle (5.5,2);
	\draw [rounded corners,fill=blue!20] (-5.2,-1) rectangle (-1.8,1);
	\draw [rounded corners,fill=blue!20] (5.2,-1) rectangle (1.8,1);
	
	\draw (0,2) node [below=4pt] {The ``open world''};
	\draw (-3.5,0) node [above=1ex] {System \Sys{}}
		node [below] {\emph{(target of attack)}};
	\draw (3.5,0) node [above=1ex] {System \Sprime{}}
		node [below] {\emph{(used to attack \Sys{})}};
		
	\draw (0,0) node [above] (A) {\parbox{20mm}{\centering Rational attacker}};
	
	\draw (-1.7,-.5) edge [<-,thick,bend right=60] (A.south);
	\draw (1.7,-.5) edge [<-,thick,bend left=60] (A.south);
	
\end{tikzpicture}
\end{center}
\caption{The attacker behaves rational outside of system $\Sys{}$'s boundaries}
\label{fig:open-world}
\end{figure}
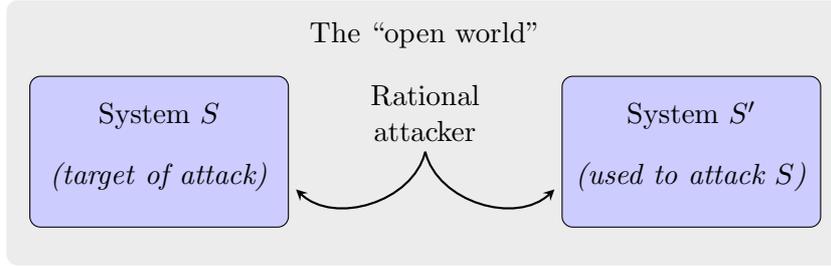

Because \Sys{} is permissionless and operating in a larger open world, \participant{} is not confined to operating exclusively within the boundaries of \Sys{}. 
\participant{} can also make use of facilities external to \Sys{}.
By assumption~2, \participant{} may in particular have access to, or be able to borrow temporarily, financial resources comparable to or larger than the total value of \Sys{}.

Suppose the facilities external to \Sys{} include another Ethereum-like cryptocurrency \Sprime{}, which includes a smart contract facility with which decentralized exchanges, futures markets, and the like may be implemented.
(This is not really a separate assumption because even if \Sprime{} did not already exist, \participant{} could create and launch it, given sufficient economic resources under assumption~2.)
Further, suppose that someone (perhaps \participant{}) has created on external system \Sprime{} a decentralized exchange, futures market, or any other mechanism by which tokens representing shares of the value of \Sys{} may be traded or speculated upon in the context of \Sprime{}: e.\,g., a series of tradeable Ethereum tokens pegged to \Sys{}'s cryptocurrency or stake units (see Fig.~\ref{fig:open-world}).

Now suppose participant \participant{} finds some behavioral strategy that system \Sys{} depends on participants \emph{not} exhibiting, and that will observably break \Sys{}---or even that just \emph{might} break \Sys{} with significant non-negligible probability.
Assumption~3 above guarantees the existence of such a behavioral strategy, unless \Sys{}'s rationality assumptions were in fact redundant and worthless.
\participant{} must merely be clever enough to find and implement such a strategy.
It is possible this strategy might first require \participant{} to pretend to be one or more well-behaved participants of \Sys{} for a while, to build up the necessary reputation or otherwise get correctly positioned in \Sys{}'s state space; a bit of patience and persistence on \participant{}'s part will satisfy this requirement.
\participant{} may also have to ``buy into'' \Sys{} enough to surmount any entry costs or stake thresholds \Sys{} might impose; the external funds \participant{} can invoke or borrow by assumption~2  can satisfy this requirement, and are bounded by the total value of \Sys{}.
In general, \Sys{}'s openness by assumption~1 and the existence of a correctness-violating strategy by assumption~3 ensures that there exists some course of action and supply of external resources by which \participant{} can position itself to violate \Sys{}'s behavioral assumption.

In addition to infiltrating and positioning itself within \Sys{}, \participant{} also invokes or borrows enough external funds and uses them to short-sell (bet against) shares of \Sys{}'s value massively in the context of the external system \Sprime{}, which (unlike \Sys{}) \participant{} trusts will remain operational and hold its value independently of \Sys{}.
Provided \participant{} reaches this short-selling position gradually and carefully enough to avoid revealing its strategy early, the funds \participant{} must invoke or borrow for this purpose must be bounded by some fraction of the total economic value of \Sys{}.
And provided there are at least some participants and/or observers of \Sys{} who believe that \Sys{} is secure and will remain operating correctly, and are willing to bet to that effect on \Sprime{}, \participant{} will eventually be able to build its short position.

Finally, once \participant{} is positioned correctly within both \Sys{} and \Sprime{}, \participant{} then launches its assumption-violating behavior in \Sys{} that will observably cause \Sys{} to fail as per assumption~2.
This might manifest as a denial-of-service attack, a correctness attack, or in any other fashion.
The only requirement is that \participant{}'s behavior creates an \emph{observable} failure, which a nontrivial number of the existing participants in \Sys{} believed would not happen because they believed in \Sys{} and its threat model.
The fact that \Sys{} is now observed to be broken, and its basic design assumptions manifestly violated, causes the shares of \Sys{}'s value to drop precipitously on external market \Sprime{}, on which \participant{} takes a handsome profit.
Perhaps \Sys{} recovers and continues, or perhaps it fails entirely---but either way, \participant{} has essentially transferred a significant fraction of system \Sys{}'s economic value from system \Sys{} itself to \participant{}'s own short-sold position on external market \Sprime{}.
And to do so, \participant{} needed only to find a way---any way---to \emph{surprise} all those who believed \Sys{} was secure and that its threat model accurately modeled \Sys{}'s real-world participants.

Even if \participant{}'s assumption-violating behavioral strategy does not break \Sys{} with perfect reliability, but only with some probability, \participant{} can still create an \emph{expectation} of positive profit from its attack by hedging its bets appropriately on \Sprime{}.
\participant{} does not need a perfect attack, but merely needs to possess the \emph{correct} knowledge that \Sys{}'s failure probability is much higher than the other participants in \Sys{} believe it to be---because only \participant{} knows that (and precisely when) it will violate \Sys{}'s design assumptions to create that higher failure probability.
Furthermore, even if \participant{}'s attack fails, and the vulnerability it exploits is quickly detected and patched, \participant{} may still profit marginally from the market's adjustment to a realization that \Sys{}'s failure probability was (even temporarily) higher than most of \Sys{}'s participants thought it was.

Within the context of system \Sys{}, \participant{}'s behavior manifests as Byzantine behavior, specifically violating the assumptions \Sys{}'s designers thought participants would not exhibit and thus excluded from \Sys{}'s threat model.
Considered in the larger context of the external world in which \Sys{} is embedded, however, including the external trading system \Sprime{}, \participant{}'s behavior is perfectly rational and economically-motivated.
Thus, the very rationality of \participant{} in the larger open world is precisely what motivates \participant{} to break, and profit from, \Sys{}'s ill-considered assumption that its participants would behave rationally.

\section{Implications for Practical Systems}

This type of financial attack is by no means entirely theoretical or limited to fully-digital systems such as cryptocurrencies.
In our scenario, \participant{} is essentially playing a game closely-analogous to the investors in
credit default swaps 
who both contributed to, and profited handsomely from,
the 2007--2008 financial crisis. 
as covered more recently in the film ``The Big Short.''

In the cryptocurrency space, some real-world attacks we are seeing---such as in\-crea\-sing\-ly-common 51\,\% attacks \cite{Attah2019}
---might be viewed as special cases of this metacircular attack on rationality. It is often claimed that large proof-of-work miners (or proof-of-stake holders) will not attempt 51\,\% attacks because doing so would undermine the value of the cryptocurrency in which they by definition hold a large stake, and hence would be ``irrational.''
But this argument falls apart if the attack allows the large stakeholder to reap rewards outside the attacked system, e.\,g., by defrauding exchanges or selling \Sys{} short in other systems.

Externally-motivated attacks on cryptocurrencies have been predicted before
in the form of
virtual protest or ``Occupy Bitcoin'' attacks~\cite{Becker2013}, 
Goldfinger attacks~\cite{Kroll2013} 
puzzle transaction attacks~\cite{Teutsch2016}, 
merged mining attacks~\cite{Judmayer2017}, 
hostile blockchain takeovers~\cite{Bonneau2018}, 
and out-of-band variants of pay-to-win attacks~\cite{Judmayer2019}. 
All these attacks are specific instances of our argument.
They have been presented in the literature as open yet solvable challenges.
We are not aware, however, of any prior attempt to summarize the lessons learned and formulate a general impossibility statement.

For most practical systems, we do not even know if they are incentive compatible in the absence of an external system \Sprime{}---i.\,e., where assumption~2 is violated---and probably they are not.
Almost all game-theoretic treatments of (parts of) the Bitcoin protocol deliver negative results.
Many attacks against specific cryptocurrency system designs are known to be profitable in expectation, such as
transaction withholding~\cite{Babaioff2012}, 
empty block mining~\cite{Houy2014}, 
selfish mining~\cite{Eyal2014}, 
block withholding~\cite{Eyal2014}, 
stubborn mining~\cite{Eyal2015}, 
fork after withholding~\cite{Know2017}, 
and
whale attacks~\cite{Liao2017} 
It is likely thanks only to frictions such as risk aversion and other costs that we rarely observe such attacks in large deployed systems.
Many specific attacks do not even depend on assumption~1, underlining the fact that rationality is not a silver bullet even where this metacircular argument does not apply. 
Where it does apply, it is more general and effectively \emph{guarantees} the existence of attacks against \emph{all} open systems that assume participants are rational.

Another related observation is that financial markets on derivatives of a system \Sys{} mature in the external world (e.\,g., \Sprime{}) as \Sys{} grows and becomes more relevant. So in some sense, systems built on the rationality assumption are temporarily more secure only until they become fat enough targets to be eaten by their own success.
We can see this effect, for example, in the growing and increasingly liquid market for hash power, which effectively thwarts Nakamoto’s~\cite{Nakamoto2008} (or Dwork's~\cite{Dwork1992}) 
rule of thumb that the ratio of processors to individuals varies in a small band.
Such dynamics happen in the real world, too. 
But there they have traditionally taken centuries or decades while in cryptocurrency space everything happens in time-lapse.

\section{Limitations of the Argument}

This argument is of course currently only a rough and informal sketch.
An enterprising student might wish to try formalizing it, or maybe someone has already done so but we are unaware of it.

The metacircular argument certainly does not apply to all cryptocurrencies or decentralized systems.
In a permissioned system, for example, in which a closed group of participants are strongly-identified and subject to legal and contractual agreements with each other, one can hope that the threat of lawsuits for arbitrarily-large damages will keep rational participants incentivized to behave correctly.
Similarly, in a national cryptocurrency, which might be relatively open but only to citizens of a given country, and which require verified identities with which the police can expect to track down and jail misbehaving participants, this metacircular argument does not necessarily apply.

Apart from police enforcement, rationality assumptions may be weakened in other ways to circumvent the metacircular argument.
For example, an open system might be designed according to a ``weak rationality'' assumption that users need incentives to join the system in the first place (e.\,g., mining rewards in Bitcoin), but that after having become stakeholders, most will then behave honestly.
In this case, rational incentives serve only as a tool for system growth, but become irrelevant and equivalent to a strong honesty assumption in terms of the internal security of the system itself.

\section{Conclusion: Irrationality Can Be Rational}

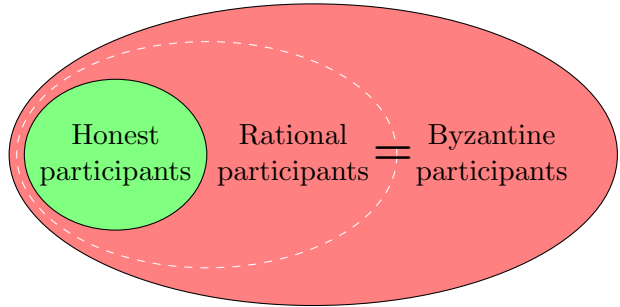
\begin{figure}[t]
\begin{center}
\begin{tikzpicture}[>=stealth,y=2cm]
	\draw [draw,fill=red!50] (0,0) ellipse (4 and 1) node [right=2.5em,black] {\parbox{25mm}{\centering Byzantine\\participants}};
	\draw [dashed,draw,white,fill=red!50] (-1.4,0) ellipse (2.5 and .75) node [right,black] {\parbox{20mm}{\centering Rational\\participants}};
	\draw (1.05,0) node [black] {\LARGE\textbf{=}};
	\draw [draw,fill=green!50] (-2.6,0) ellipse (1.2 and .5) node [black] {\parbox{20mm}{\centering Honest\\participants}};
\end{tikzpicture}
\end{center}
\caption{Rational and Byzantine participants are equivalent in permissionless systems}
\label{fig:adversaries-2}
\end{figure}

What many in the cryptocurrency community seem to want is a system that is both permissionless
and tolerant of strongly-rational behavior---either beyond the thresholds a similar a Byzantine system would tolerate
(such as a rational majority), or by deriving some simplicity or efficiency benefit from assuming rationality.
But in an open world in which the permissionless system is not the only game in town, a potential \emph{perfectly rational} attacker can always exist, or appear at any time, whose entirely rational behavior is precisely to profit from bringing the system down by violating its assumptions on participant behavior.

So if you think you have designed a permissionless decentralized system that is cleverly secured based on rationality assumptions, you haven't. You have merely obfuscated the rational attacker's motive and opportunity to profit outside your system from breaking your rationality assumptions. The only practical way to eliminate this threat appears to be either to close the system and require strong identities and police protection, or else secure the system against arbitrary Byzantine behavior, thereby rendering rationality assumptions redundant and useless for security (cf.~Fig.~\ref{fig:adversaries-2}).

\vfill
\subsubsection*{Acknowledgements}
\small We wish to thank Jeff Allen, Ittay Eyal, Damir Filipovic, Patrik Keller, Alexander Lipton, Andrew Miller, and Haoqian Zhang for helpful feedback on earlier drafts of this work.

\bibliographystyle{plain}
\bibliography{rationality}

\end{document}